\documentstyle[aps,twocolumn]{revtex}
\begin{document}
\draft

\title{Stability Exponents, Separation of Variables, and Lyapunov Transforms}

\author{ William E. Wiesel}
\address{ Air Force Institute of Technology/ENY \\
 2950 P Street \\
 Wright--Patterson AFB, Ohio 45433  } 
\maketitle

\begin{abstract}
   The problem of formulating self-consistent local and global
stability exponents is shown to require global separation of
variables.  Posing the separation of variable problem, we see
that many such separations are possible, but only one is 
consistent with both Hamiltonian dynamics and the boundedness
requirement for a Lyapunov transform: the determinant of the
modal matrix must be constant.  Such stability exponents are
invariant to any linear transformation of variables, and both
the local stability exponents and modal matrix appear to be
point functions in the original space, and introduce a true
coordinate frame.  Methods are presented to perform this
separation at equlibrium points, about periodic orbits, and
along general trajectories.  Results of numerical experiments
are given.
\end{abstract}
\narrowtext

\input epsf.sty

\section{Introduction}

     A nonlinear dynamical system can be written as
\begin{equation}
\dot {\bf X} = {\bf f}( {\bf X} ) .
\label{nonlinear}
\end{equation}
We shall consider only autonomous dynamical systems in this paper.  
If the system itself is time dependent, then defining an 
additional state variable $ x_{N+1} = t $ and appending
$ \dot x_{N+1} = 1 $ to (\ref{nonlinear}) will render the
system autonomous.  
Writing $ {\bf x} = {\bf X}(t) - {\bf X}_0 (t) $ as the
difference between a nominal solution $ {\bf X}_0 $ and a
nearby trajectory, to first order $ {\bf x} $ obeys
\begin{equation}
\dot {\bf x} = \left. {{ \partial {\bf f}} \over {\partial 
  {\bf X}}} \right|_{ {\bf X}_0 } {\bf x} 
 =  A(t) {\bf x} .
\label{smallx}
\end{equation}
As a linear system, the complete solution to the above is
contained within the fundamental matrix $ \Phi $, which
obeys
\begin{equation}
\dot \Phi = A(t) \Phi, \;\;\; \Phi(t_0 ) = I .
\label{varyeqns}
\end{equation}
At least for finite final times, the fundamental matrix can
be constructed by numerical integration.
The general solution to (\ref{smallx}) is then written
as
\begin{equation}
{\bf x}(t) = \Phi(t, t_0 ) {\bf x}(t_0 ) ,
\label{smalldisp}
\end{equation}
where the two time indices on $ \Phi $ indicate the end and
start times, respectively.

     The stability and control of chaotic systems are usually
discussed in terms of Lyapunov exponents.  As is well known, 
Lorenz \cite{lor84}, Greene and Kim \cite{gre87}, Lyapunov exponents
appear in the singular value / singular vector decomposition of
the fundamental matrix
\begin{equation}
\Phi( t_f, t_0 ) = U \exp( \Lambda ( t_f - t_0 )) V^T ,
\label{singval}
\end{equation}
where $ U $ and $ V $ are real orthonormal matrices, and
where the elements of the real diagonal matrix $ \Lambda $ are
the Lyapunov exponents.  As $ t_f \rightarrow \infty$, the
orthonormal matrix $ V $ approaches a constant, and a 
local coordinate frame is introduced near the trajectory.

     But this decomposition does not match the usual decompositions
for constant coefficient and periodic coefficient systems.  For
constant coefficient systems, the usual decomposition is
\begin{equation}
\Phi( t_f , t_0 ) = E \exp( \Lambda (t_f - t_0 )) E^{-1} ,
\label{conscoef}
\end{equation}
where $ E $ is the matrix of eigenvectors and $ \Lambda $ is the
Jordan normal form containing the eigenvalues of the
constant matrix $ A $.  This form is only possible because 
$ A $ is constant.  In the periodic coefficient case, Floquet
showed that the fundamental matrix decomposed as
\begin{equation}
\Phi( t, t_0 ) = E(t) \exp( \Lambda ( t - t_0 )) E^{-1} (t_0 ) .
\label{periodcoef}
\end{equation}
Here, the matrix $ E(t) $ is time dependent but periodic, while
the Jordan form $ \Lambda $ contains the Poincar\'e exponents.  The
most obvious difference between (\ref{conscoef}), (\ref{periodcoef}) 
and (\ref{singval}) is that the former two permit imaginary
parts in the stability exponents, while Lyapunov exponents are
purely real quantities.

    There are more subtle differences as well between Lyapunov
exponents and the other stability exponents.  The Floquet 
decomposition, when applied to an equilibrium point immediately
reduces to the usual constant coefficient solution.  The 
decomposition (\ref{singval}) does not reduce to either of the
other two cases.   Also, the state vector $ {\bf X}$ 
usually consists of disparate physical
quantities, often not measured in the same physical units.  Assume
two realizations of the state vector are $ {\bf X} $ and $ {\bf Y}$, 
related by the coordinate transformation law
\begin{equation}
{\bf X} = {\cal X} ( {\bf Y} )  .
\end{equation}
Then two representations of the local state vector are related by
\begin{equation}
{\bf x}(t) = M(t) {\bf y}(t) ,
\label{lcljacobian}
\end{equation}
where $ M = \partial {\cal X} / \partial {\bf Y} $ 
is the Jacobian matrix of the transformation of coordinates.
[The Jacobian is actually a function of position $ {\bf X}$, 
but we use the reference trajectory $ {\bf X}_0 (t) $ to write it as
a function of time.  This is, however, only a convenient
shorthand.  We are considering autonomous coordinate 
transformations, and $ M $ is not an explicit function
of time.]  Using the
above in (\ref{smalldisp}), and remembering that the fundamental matrix
is the derivative of the final state with respect to the
initial state, 
$ \Phi_y (t, t_0 ) = \partial {\bf y}(t) / \partial {\bf y} (t_0 )$, 
the fundamental matrix expressed in the variables $ {\bf y} $ is
\begin{equation}
\Phi_y (t_f , t_0 ) = M^{-1}(t_f) \Phi_x (t_f, t_0 ) M(t_0)  .
\label{phixform}
\end{equation}
Inspection of (\ref{conscoef}) and (\ref{periodcoef}) shows that
these forms are compatible with this transformation.  In particular, 
the matrix $ \Lambda $ of stability exponents is invariant to any 
autonomous change in coordinate systems.  
This is true in the constant coefficient case since the reference
trajectory is a point, and so $ M(t_f) = M(t_0 )$.  This makes
$ \Phi_y $ similar to $ \Phi_x $.  In the periodic case, the
stability exponents are evaluated from the $ \Phi $ matrix 
at one full period, and so again the Jacobian matrices are
the same, and the $ \Phi $ matrices are similar.  Stability
exponents in these cases are fully invariant to \it any \rm autonomous
coordinate transformation.

     On the other hand, the classical Lyapunov exponents 
(\ref{singval}) are only invariant
under \it rotations \rm of the original coordinate system.  
They are not even invariant under changes in physical units.  
This is a nontrivial matter when one is not working in an abstract 
``metric space'', but instead has to contend with physical units
for a physical problem.  

    Also, the ``local'' Lyapunov exponents deserve mention
here, e.g. \cite{aba91}, \cite{eck93}.  They are the 
eigenvalues of $ A^T A $ at a point, and 
are an attempt to define a quantity which would characterize
the instantaneous growth and decay rates of small displacements.
However, since they ignore terms relating to the rate of change
of the eigenvectors of $ A^T A $, they do not integrate with time
to give the global Lyapunov exponents.

    In earlier works we have explored extensions of the constant
coefficient and periodic coefficient decompositions to the general
case of (\ref{smallx}).  In Wiesel \cite{wie92}, we used the
decomposition 
$ \Phi(t, t_0 ) = E(t) \exp( \Lambda(t)(t - t_0 )) E^{-1}(t)$,
which is just the instantaneous eigenvalue decomposition.  But
the stability exponents of this form will not be invariant
to most coordinate transformations either, and the assumption
that $ E(t) = E(t_o )$ seems unnatural.  In 
\cite{wie93} the decomposition (\ref{periodcoef}) was extended
to more general systems, and winding numbers were successfully
calculated as the imaginary parts of 
the constant matrix $\Lambda$.  But
that paper only investigated deterministic regions of conservative
Hamiltonian systems.  In this paper we will explore the conditions
under which the invariance of the stability exponents of the
constant coefficient case can be
extended, and the extent to which self--consistent local and
global stability exponents can be defined.

\section{Lyapunov Transforms}

     Consider a slight extension of the Floquet decomposition
(\ref{periodcoef}) to
\begin{equation}
\Phi (t_f, t_0 ) = E(t) \exp ( \Omega( t) ) E^{-1}(t_0 ) .
\label{mydcmp}
\end{equation}
The matrix $  \Omega $ will be presumed diagonal, since it must
commute with its derivative for what is to follow.  
Stability exponents over the time interval then are
\begin{equation}
\omega_i = {{ \Omega_{ii} (t_f) } \over { t_f - t_0 }} .
\label{exponents}
\end{equation}
Inserting (\ref{mydcmp}) into (\ref{varyeqns}) and rearranging
produces
\begin{equation}
\dot E = A E - E \dot \Omega  ,
\label{Edot}
\end{equation}
which assumes that $ \dot \Omega $ and $ \Omega(t) $ commute.  
If the matrix determinant of $ E(t) $ is bounded from both above and
below, 
\begin{equation}
\left| \left| E(t) \right| \right| \leq \rho, \;\;\;
\left| \left| E^{-1}(t) \right| \right| \leq \rho, 
\label{bound}
\end{equation}
for all $ t $ within $ t_0 \leq t \leq t_f $ 
then (\ref{mydcmp}) is a \it Lyapunov transformation\rm, 
Rugh \cite{rug93}.  That is, it is an example of the most general
type of linear transformation of (\ref{smallx}) that will preserve
stability information.   We will refer to the $ \omega_i $ [and
their limits 
as $ t_f \rightarrow \infty $] as \it extended \rm stability
exponents.  They may have imaginary parts.  

     Without the conditions on the determinant of $ E(t) $, the transformation
(\ref{mydcmp}) is very general.  If we specify \it any \rm diagonal
matrix function $ \Omega(t) $ and choose a 
\it random \rm $ E(t_0) $, then (\ref{Edot}) will produce an $E(t_f)$  
which will reproduce $ \Phi (t_f, t_0 ) $ through (\ref{mydcmp}).  
In most cases the determinant of $ E$ will collapse or expand exponentially.  
This is not desirable, since this matrix can be interpreted as a
local coordinate transformation.  Let
\begin{equation}
{\bf x} = E(t) {\bf y}  .
\label{coordxform}
\end{equation}
Then calculating $ \dot {\bf x}$, and using (\ref{smallx}) and
(\ref{Edot}), we find
\begin{equation}
\dot {\bf y} = \dot \Omega {\bf y}  .
\label{modaleom}
\end{equation}
That is, the transformation (\ref{coordxform}) 
will decouple the equations
of motion of the linear system.  This decoupling is only meaningful, 
however, if $ E $ and its inverse are well behaved, so that the
coordinate transformation (\ref{coordxform}) is legitimate.  

    It is interesting, however, to compare (\ref{modaleom}) to
the analogous form (\ref{smallx}).   Suppose that we calculated
the variational equations (\ref{smallx}) for a dynamical 
system, and found that they were of the form (\ref{modaleom})?
Then (\ref{modaleom}) would not be a local diagonalization
along one particular trajectory, but would diagonalize all
trajectories simultaneously.  If the new form (\ref{modaleom})
is \it globally \rm the variational 
equations of a transformed version of the
original dynamical system, then the system that gives rise to
(\ref{modaleom}) must have the form
\begin{equation}
\dot {\bf Y}_i = {\bf g}_i ({\bf Y}_i).
\label{decoupledeom}
\end{equation}
The important thing to notice here is that the above equations,
unlike (\ref{nonlinear}), are \it decoupled\rm.  We now reverse
the entire chain of deduction that led to this point.  In order to
produce local stability exponents $ \dot \Omega $ which integrate
along \it any \rm trajectory 
to give global stability exponents $ \Omega$, the
dynamical system must be \it decoupled\rm.
The existence of a global
Lyapunov transformation implies such a decoupling, and vice
versa.

\section{Separation Near an Equlibrium}

    In this section we will concentrate on
the vicinity of equlibrium points.  Eq (\ref{coordxform}) then
becomes the linear portion of a series expansion about an
equlibrium
\begin{eqnarray}
 x_i &=& E_{i \alpha} y_{\alpha} + {1 \over {2!}} 
{{ \partial E_{i \alpha}} \over {\partial y_{\beta}} }
y_{\alpha} y_{\beta} \nonumber \\
&+& {1 \over {3!}} {{ \partial^2 E_{i \alpha}} \over
{ \partial y_{\beta} \partial y_{\gamma} }} y_{\alpha}
y_{\beta} y_{\gamma} + ...    .
\label{coordseries}
\end{eqnarray}
[Greek indices indicate summations, while roman indices
are not summed.  Also, instead of noting that
these quantities are evaluated at an equlibrium, we will find it
more convenient to explicitly show functional dependence wherever
a quantity is \it not \rm evaluated at the equlibrium.]  
We note that not all such series expansions will yield true
coordinate frames.  The condition that the new $ {\bf y} $
be a coordinate frame is symmetry in all indices except the
first:
\begin{equation}
{{\partial E_{i \alpha}} \over {\partial y_{\beta}}} = 
{{\partial E_{i \beta}} \over { \partial y_{\alpha}}},  \;\;\;
{{\partial^2 E_{i \alpha}} \over {\partial y_{\beta} \partial y_{\gamma}}}=
{{\partial^2 E_{i \beta}} \over {\partial y_{\alpha} \partial y_{\gamma}}}=
{{\partial^2 E_{i \gamma}} \over {\partial y_{\beta} \partial y_{\alpha}}},
\label{coordexpansion}
\end{equation}
and so forth.  [The familiar Lie bracket conditions apply to the covariant 
derivatives, not the contravariant quantities above.]  
Along with this expansion, we have the
parallel expansion of the new equations of motion
about the equlibrium
\begin{equation}
g_i = \dot \Omega_i y_i + {1 \over {2!}}
{{\partial \dot \Omega_i } \over {\partial y_i}} y_i^2
+ {1 \over { 3!}} {{\partial^2 \dot \Omega_i } \over { \partial y_i^2 }}
y_i^3 + ... \;\; .
\label{eomexpansion}
\end{equation}
Separation of variables mandates that each $ \dot \Omega_i $ be a
function \it only \rm of the corresponding $ y_i $.  As an
immediate corollary, each local stability exponent $ \dot \Omega_i $
is constant on surfaces of constant $ y_i $.

     We begin by noting in a general coordinate transformation
$ {\bf X} = {\bf X}({\bf Y})$ that 
\begin{equation}
\dot {\bf Y}= {\bf g}  = E^{-1}({\bf Y}) {\bf f} ({\bf X}({\bf Y})) ,  
\label{capYdot}
\end{equation}
where $ E = \partial {\bf X} / \partial {\bf Y} $ is the
Jacobian of the transformation.  At the equlibrium point, then,
$ g_i = 0$.  Proceeding to the first order, we have
\begin{eqnarray}
{{ \partial {\bf g}_i } \over { \partial y_j}} &=& 
   A_{Y,ij} \\
&=& E^{-1}_{i \alpha} {{\partial f_{\alpha}} \over {\partial y_j }}
+ {{ \partial E^{-1}_{i \alpha }} \over {\partial y_j}} 
f_{\alpha} \nonumber \\
&=& E^{-1}_{i \alpha} {{\partial f_{\alpha}} \over {\partial x_{\beta}}}
E_{\beta j} - E^{-1}_{i \beta} {{\partial E_{\beta \gamma}} \over
{\partial y_j }} E^{-1}_{\gamma \alpha} f_{\alpha} \nonumber \\
&=& E^{-1}_{i \alpha} A_{X,\alpha \beta} E_{\beta j} 
 - E^{-1}_{i \beta} {{\partial E_{\beta \gamma}} \over
{\partial y_j }} E^{-1}_{\gamma \alpha} f_{\alpha} . \nonumber 
\end{eqnarray}
We have replaced the $ y $ partial of $ {\bf f} $ with its equivalent
in terms of $ x$, recognized $ A_X = \partial {\bf f} / \partial x$, 
and expanded the derivative of the inverse matrix.  Then, evaluating
at the equlibrium point, we must have
\begin{equation}
A_{Y,ij} = \dot \Omega = E^{-1}_{i \alpha} A_{ X, \alpha \beta}
E_{\beta j}  .
\label{order1}
\end{equation}
To effect a separation of variables this must be diagonal, and
we are immediately forced into using the eigenvalue / eigenvector
decomposition of $ A_X $ as the first order values for $ \dot \Omega_i $
and $ E $.

   At the second order, we calculate the second partial derivative
of (\ref{capYdot}), and evaluate it at the equlibrium.  The 
result is
\begin{eqnarray}
{{\partial^2 g_i } \over {\partial y_j \partial y_k}} &=&
 E^{-1}_{i \alpha} \left\{ 
 {{ \partial E_{\alpha j}} \over { \partial y_k }} \dot \Omega_i
-{{ \partial E_{\alpha k}} \over { \partial y_j }} \dot \Omega_k
-{{ \partial E_{\alpha j}} \over { \partial y_k }} \dot \Omega_j
\right. \nonumber \\
&+& \left. {{ \partial A_{X,\alpha \beta}} \over {\partial x_{\gamma}}}
E_{\gamma k} E_{\beta j} \right\} .
\label{order2}
\end{eqnarray}
Again, for a separation of variables we must require that
the above expression be zero, except for $ i = j = k$, when
it must equal $ \partial \dot \Omega_i / \partial y_i $.  Note that
(\ref{order2}) constitutes $ N^3 $ linear equations in the
$ N^3 $ unknowns $ \partial E_{ij} / \partial y_k $, but with
the $ N $ additional unknown quantities 
$ \partial^2 g_i / \partial y_i^2 = \partial \dot \Omega_i / \partial y_i$.  
For the moment we delay specifying additional information to
choose the quantities $ \partial \dot \Omega_i / \partial y_i $.

    What we are attempting is similar to center manifold theory, 
and reduces to it if we drop all of the ``diagonal'' terms from
(\ref{order2}), e.g. Arrowsmith and Place \cite{arr90}.
In center manifold theory only the
surfaces containing the trajectories are obtained.  Our aim in
this work goes beyond just obtaining the manifolds through
the equlibrium.  Also, this approach is similar to 
normal form theory. 
In fact, if we chose all the derivatives of $ \dot \Omega_i $
to be zero, we would be attempting to map the entire dynamical
system back onto the equlibrium point variables, and this \it would \rm
be normal form theory.  This is also \it not \rm our aim.

     Continuing to the third order we obtain
\begin{eqnarray}
& &{{\partial^3 g_i } \over { \partial y_j \partial y_k \partial y_l }} =
E^{-1}_{i \alpha} {{ \partial^3 f_{\alpha} } \over { \partial y_j
\partial y_k \partial y_l }} \label{order3} \\
&+&
{{ \partial E^{-1}_{i \alpha} } \over { \partial y_j }}
{{\partial^2 f_{\alpha} } \over { \partial y_k \partial y_l }} +
{{ \partial E^{-1}_{i \alpha} } \over { \partial y_k }}
{{\partial^2 f_{\alpha} } \over { \partial y_j \partial y_l }} +
{{ \partial E^{-1}_{i \alpha} } \over { \partial y_l }}
{{\partial^2 f_{\alpha} } \over { \partial y_j \partial y_k }}
\nonumber \\
&+&
{{ \partial^2 E^{-1}_{i \alpha} } \over { \partial  y_j  \partial y_k }}
{{ \partial f_{\alpha} } \over {\partial y_l }} +
{{ \partial^2 E^{-1}_{i \alpha} } \over { \partial  y_j  \partial y_l }}
{{ \partial f_{\alpha} } \over {\partial y_k }} +
{{ \partial^2 E^{-1}_{i \alpha} } \over { \partial  y_k  \partial y_l }}
{{ \partial f_{\alpha} } \over {\partial y_j }} , \nonumber
\end{eqnarray}
where
\begin{equation}
{{\partial f_{i} } \over { \partial y_j }} = A_{i \beta}
E_{\beta j} ,
\end{equation}
\begin{equation}
{{\partial^2 f_{i} } \over { \partial y_j \partial y_k }} = 
{{\partial A_{i \beta } } \over { \partial x_{\gamma} }}
E_{\gamma k} E_{\beta j} + A_{i \beta} 
{{\partial E_{\beta j} } \over { \partial y_k }} ,
\end{equation}
\begin{eqnarray}
& & {{ \partial^3 f_i } \over { \partial y_j \partial y_k 
\partial y_l }} = {{\partial^2 A_{i \beta} } \over 
{ \partial x_{\gamma} \partial x_{\delta} }} E_{\delta l}
E_{\gamma k} E_{\beta j} \nonumber \\
&+&
{{\partial A_{i \beta} } \over { \partial x_{\gamma} }} \left(
{{\partial E_{\gamma k} } \over {\partial y_l }} E_{\beta j} +
{{\partial E_{\beta j} } \over { \partial y_l }} E_{\gamma k} +
{{\partial E_{\beta j} } \over { \partial y_k }} E_{\gamma l} \right)
\nonumber \\
&+&
A_{i \beta} {{\partial^2 E_{\beta j} } \over { \partial y_k
\partial y_l }} ,
\end{eqnarray}
and
\begin{equation}
{{\partial E^{-1}_{ik} } \over { \partial y_j }} =
- E^{-1}_{i \beta} {{\partial E_{\beta \sigma} } \over { \partial y_j }}
E^{-1}_{\sigma k}  ,
\end{equation}
\begin{eqnarray}
& & {{\partial^2 E^{-1}_{ij} } \over { \partial y_k \partial y_l }} = 
- {{\partial E_{i \beta} } \over { \partial y_l }}
{{\partial E_{\beta \sigma} } \over { \partial y_k }} E^{-1}_{\sigma j}
\nonumber \\
&-&
E^{-1}_{i \beta} {{\partial^2 E_{\beta \sigma} } \over { \partial y_k
\partial y_l }} E^{-1}_{\sigma j} - 
E^{-1}_{i \beta} {{\partial E_{\beta \sigma} } \over { \partial y_k }}
{{\partial E^{-1}_{\sigma j} } \over { \partial y_l }} .
\end{eqnarray}

     Again, to force separation of variables, (\ref{order3}) must
equal zero except for the $ N$ cases where $ i=j=k=l$, in
which case it equals $ \partial^2 \dot \Omega_i / \partial y_i^2$.
These are thus $ N^4 $ linear equations in the $ N^4 + N$
unknowns $ \partial^2 E_{ij} / \partial y_k \partial y_l $
and $ \partial^2 \dot \Omega_i / \partial y_i^2 $.

    Before proceeding to methods to specify the extra variables
at each order, we will first establish some properties of this
transformation.  First, 
$ g_i = E^{-1}_{i,\alpha}({\bf Y}) {\bf f}_{\alpha}({\bf X}({\bf Y})) $
is presumably a continuous, differentiable function of $ {\bf Y}$, 
and therefore 
$ \partial g_i / \partial y_j \partial y_k = \partial g_i /
\partial y_k \partial y_j$.  Inserting (\ref{order2}) into
this expression and using the symmetry of derivatives of $ A$, 
three terms immediately cancel, leaving
\begin{equation}
E^{-1}_{i \alpha} {{\partial E_{\alpha j}} \over {\partial y_k}}
\dot \Omega_i = E^{-1}_{i \alpha} {{\partial E_{\alpha k}} \over
{\partial y_j}} \dot \Omega_i ,
\end{equation}
and a simple further simplification confirms the symmetry of
$ \partial E_{\alpha j} / \partial y_k $.  This is required if
the new variables $ {\bf Y} $ are to form a coordinate frame.
At the third order, extensive numerical calculations have 
failed to produce a non-symmetric 
$ \partial^2 E_{ij} / \partial y_k \partial y_l $.  

    To see why (\ref{order2}) and (\ref{order3}) leave the 
stability exponent unspecified, consider the simple dynamical 
system
\begin{eqnarray}
\dot x &=& {x \over { \sqrt{ x^2 + y^2 }}} - x - y , \nonumber \\
\dot y &=& {y \over { \sqrt{ x^2 + y^2 }}} + x - y .
\end{eqnarray}
This is the rectangular form of the polar variable differential
equations
\begin{equation}
\dot r = 1 - r, \;\;\;  \dot \theta = 1 ,
\end{equation}
so this system obviously separates in polar coordinates
$ Y_1 = r$, $ Y_2 = \theta$.  But it \it also \rm separates in 
any coordinate frame $ {\cal Y} $ which itself
is a separable function of the polar coordinates
\begin{equation}
{\cal Y}_1 = h_1 ( r), \;\;\; {\cal Y}_2 = h_2 ( \theta ) .
\end{equation}
There is thus an infinite number of coordinate frames in which a
separable system can be separated.  The undetermined stability
exponent derivatives in (\ref{order2}) and (\ref{order3}) are
\it directly \rm related to the derivatives of the arbitrary
functions $ h_i $ above.

      So, if the
system separates in terms of the variables $ {\bf Y}$, then
it also separates in any variables $ {\cal Y}_i = {\cal Y}_i ( Y_i)$,
where each $ {\cal Y}_i $ is an arbitrary function of one $ Y_i $.
The equations of motion in terms of these new variables 
take the form
\begin{equation}
\dot {\cal Y}_i = \left( {{ dY_i } \over { d {\cal Y}_i }} 
( {\cal Y}) \right)^{-1}
g_i ( Y_i ( {\cal Y}_i )) .
\end{equation}
The local stability exponents for the new variables are 
$ \dot \Omega_{i, {\cal Y}} = \partial \dot {\cal Y}_i / \partial 
{\cal Y}_i$, and direct calculation yields
\begin{equation}
\dot \Omega_{i, {\cal Y}} = \dot \Omega_{i, Y} -
 {{ d^2 Y_i } \over { d{\cal Y}_i^2 }} ( {\cal Y})
 g_i( Y_i ({\cal Y}_i ))  
 \left( {{ dY_i } \over { d{\cal Y}_i }} 
 ({\cal Y}) \right)^{-2} .
\end{equation}
This result shows that these stability exponents are invariant
to constant coefficient linear transformations, where 
the second derivative is zero.  
If we study the same equlibrium beginning in two different
coordinate frames, we will obtain the same first order stability
exponents, and the eigenvectors will have the same direction, 
but almost inevitably different magnitudes.  The above result, 
however, shows that the higher order stability exponent derivatives
will yield the same exponents when evaluated at the same 
point $ {\bf X} $ in physical space.  This result has also
been confirmed numerically.  It implies that 
$ \dot \Omega_i ({\bf X}) $ is a true point function of
the original space position vector $ {\bf X}$.  [This is not
true of $ \Omega$, which for finite times will be a function
of the arc studied.]  This invariance class is also significantly
stronger than the usual Lyapunov exponents, which are only
invariant under rigid rotations of the original coordinate frame.

     To ensure that the bound conditions (\ref{bound}) are met, 
and that the coordinate transformation (\ref{coordxform}) is
nonsingular, we investigate the determinant of $ E(t)$.  
Calculating the derivative of the determinant and
using (\ref{Edot}), one obtains
\begin{eqnarray}
{d \over {dt}} \left| E(t) \right| &=&
\sum_{i=1}^N \left| {\bf e}_1 ... \dot {\bf e}_i
  ... {\bf e}_N \right| \\
&=& \sum_{i=1}^N \left| {\bf e}_1 ... A{\bf e}_i ... {\bf e}_N \right|
 - \sum_{i=1}^N \dot \Omega_i \left| E(t) \right|  .  \nonumber
\end{eqnarray}
The first term can then be reduced by a standard argument \cite{mei70} to
produce
\begin{equation}
{d \over {dt}} \left| E(t) \right| = \left( {\rm Tr}( A(t)) - 
{\rm Tr}( \dot \Omega) \right) \left| E(t) \right|  ,
\label{detEdot}
\end{equation}
where $ {\rm Tr}() $ is the trace.  This has solution
\begin{equation}
\left| E(t) \right| = \left| E(t_0 ) \right| \exp \left\{
\int_{t_0}^t \left( {\rm Tr}( A(t)) - 
{\rm Tr}( \dot \Omega(t)) \right) dt \right\}  .
\end{equation}
This form directly shows that $ \left| E(t) \right| $ cannot
become zero, so a lower bound exists over any finite time interval.  
Remembering that the $ \dot \Omega_i $ are local stability
exponents, to obtain an upper bound in the long term it is
most consistent to impose the instantaneous condition
\begin{equation}
{\rm Tr} \dot \Omega = {\rm Tr} A  .
\label{constantTr}
\end{equation}
This condition will ensure that $ \left| E(t) \right| $
is constant for all time.

      In the case of Hamiltonian systems, 
we would wish that (\ref{coordseries}) be a canonical transform.  
Since the determinant of the Jacobian of any 
canonical transform (e.g. $ E $ here)
must be +1 everywhere, we are led to specify that 
$ \left| E \right| $ is constant.  [It is also required that
$ E $ be symplectic, but it is well known that (\ref{Edot})
stays symplectic if $ E (t_0 )$ is symplectic \cite{wie94}.]  
Therefore, Hamiltonian systems demand that $ \left | E(t) \right| $
be constant.  In the case of dissipative systems, we still 
must face the boundedness requirements (\ref{bound}) on 
Lyapunov transforms.  If (\ref{constantTr}) is true, then
$ \left| E(t_f ) \right| = \left| E(t_0 ) \right| $, and
since under reasonable assumptions (\ref{detEdot}) is continuous
and bounded, then $ \left| E(t) \right| $ is bounded away from
infinity.  Finally, there is the practical point that 
it is hard to program ``boundedness'' in an algorithm, but 
relatively easy to program constancy.   We will make this
latter choice, and enforce (\ref{constantTr}).

     It is now possible to explicitly determine all of the 
partial derivatives of the stability exponents about an
equlibrium point.  Beginning
with (\ref{constantTr}), we remember that each $ \Omega_i $
is a function only of $ y_i $ to find
\begin{eqnarray}
{\partial \over {\partial y_i}} {\rm Tr} \dot \Omega &=&
 {{ \partial \dot \Omega_i } \over { \partial y_i }} 
 = {{ \partial A_{\alpha \alpha} } \over { \partial y_i }} \nonumber \\
 &=& {{ \partial A_{\alpha \alpha} } \over { \partial x_{\beta}} }
 E_{\beta i} .
\label{dOmdy}
\end{eqnarray}
It is remarkable that one constraint (\ref{constantTr}) combined
with the separation condition determines all of the local
stability exponents individually.  We also note that the
above form preserves the usual constant--coefficient case:
$ \partial \dot \Omega_i / \partial y_i = 0 $ if the system
really is a constant coefficient linear problem,
$ \partial A_{ij} / \partial x_k = 0$.
Similarly, at the next order we have
\begin{eqnarray}
{{\partial^2 \dot \Omega_i } \over { \partial y_i^2 }} &=&
{{ \partial^2 A_{\alpha \alpha} } \over { \partial y_i^2 }} \nonumber \\
&=& {{\partial^2 A_{\alpha \alpha} } \over { \partial x_{\beta} 
\partial x_{\gamma}} } E_{\beta i} E_{\gamma i} 
 + {{\partial A_{\alpha \alpha} } \over {\partial x_{\beta}}} 
 {{\partial E_{\beta i}} \over {\partial y_i}}.
\end{eqnarray}
This immediately generalizes to any order.  Knowing the partial
derivatives of $ \dot \Omega$, it is then generally possible to
solve (\ref{order2}), (\ref{order3}), and subsequent systems
of linear equations for the coefficients of the 
coordinate frame transformation.  Actually, we have used
(\ref{order2}) and (\ref{order3}) for numerical checks.
Appendix A presents an alternate form of these conditions
that is considerably more efficient numerically.  

     The one consistent case where a solution is not possible
is where the zero order stability exponents occur as 
positive / negative pairs.  Such equlibrium points (including
all Hamiltonian equilibria) produce a singular matrix at
the third order in (\ref{order3}).  The question of the
existence of separation transformations near centers and
saddles will be investigated by other means later in this
paper.

    Other choices for the stability exponents are possible, and
we have investigated some of them.  It is possible to extremalize
$ \left| \partial \dot \Omega_i / \partial y_i \right|^2 $
subject to the separation conditions (\ref{order2}), in an
attempt to pick maximal stability exponents.  Unfortunately, the
result is always zero for the stability exponent partial
derivatives, which certainly is an extremal answer.  Norm--like
quantities can also be extremalized subject to the separation
conditions, for example
$ \sum_{ijk} \left( \partial E_{ij} / \partial y_k \right)^2 $
is one such quantity.  We also note that in structural mechanics there
is a theory of ``nonlinear normal modes'', Vakakis et. al
\cite{vak96}, which minimizes the curvature of the new coordinate
frame.  We have not elected that approach here.  Curvature
can be uniquely defined in the theory of structures, in general
relativity, and in differential geometry where an underlying
``flat'' space is implicitly assumed.  But in general dynamical
systems it is traditional to use \it any \rm set of acceptable
coordinates for a system, and there may be no good answer to
the question of which set of original coordinates is the
``flat'' one.

    The author believes that the choice of constant determinant
of the Jacobian matrix $ E $ is compelling.  It is the only
choice for Hamiltonian systems which makes it possible to
have a canonical transformation $ {\bf X} \rightarrow {\bf Y}$.
For non-Hamiltonian systems other choices may exist which
bound $ \left| E \right|$, but the author is unaware of any
easily specified transformation which enforces this essential
requirement of the transformation.

     Since Hamiltonian systems form a good part of the reasons
for the choice of the local stability exponents, it is perhaps
not surprising that they assume a special form in this case.
Partition the state vector of a $ 2N$ order system with
Hamiltonian $ H $
as $ {\bf X}^T = \left\{ q_i ,\; p_i \right\}$.  Then direct
calculation gives
\begin{equation}
A_{ii} = {{ \partial^2 H } \over { \partial p_i \partial q_i }},
\;\;\;
A_{i+N,i+N} = - {{ \partial^2 H } \over {\partial p_i \partial q_i }} .
\end{equation}
Continuing, the local stability exponents are given by, in 
explicit summation notation
\begin{eqnarray}
{{\partial \dot \Omega_i } \over {\partial y_i }} &=&
\sum_{\alpha=1}^N \sum_{\beta=1}^{2N} \left\{
{{ \partial^3 H } \over {\partial p_i \partial q_i \partial x_{\beta} }}
E_{\beta i} -
{{ \partial^3 H } \over {\partial p_i \partial q_i \partial x_{\beta} }}
E_{\beta i} \right\} \nonumber \\
&\equiv& 0  \label{HamdOmdy}
\end{eqnarray}
for Hamiltonian systems.

\section{Separation Along a Trajectory}

    The range of the equlibrium point expansion can be
extended by sampling the solution space along trajectories
that emanate from the equlibrium.  Returning to the differential
equation forms, we can simultaneously integrate the equations
of motion (\ref{nonlinear}), the modal differential equations
(\ref{Edot}), and
\begin{equation}
{d \over {dt}} \Omega_i = \dot \Omega_i .
\label{Omegadot}
\end{equation}
To obtain a complete set of ordinary differential
equations, it is necessary to also produce a
differential equation for $ \ddot \Omega_i$.  Returning 
to (\ref{dOmdy}), and
remembering that each $ \dot \Omega_i $ is a function of only
one $ y_i$, we find
\begin{eqnarray}
{ d \over {dt}} \dot \Omega_i &=& {{ \partial \dot \Omega_i } \over
{ \partial y_i }} \dot y_i 
= {{ \partial A_{\alpha \alpha} } \over { \partial x_{\beta} }}
E_{\beta i} \dot y_i \nonumber \\
&=& {{\partial A_{\alpha \alpha} } \over { \partial x_{\beta} }}
E_{ \beta i} E^{-1}_{i \gamma} f_{\gamma} .
\label{Omegadotdot}
\end{eqnarray}
At this point we have a closed set of differential equations.  
In addition, 
\begin{equation}
 \dot Y_i = E^{-1}_{i \alpha} f_{\alpha} 
\label{ydot}
\end{equation}
can also be integrated to map the original coordinate frame
$ {\bf X} $ onto the new frame $ {\bf Y}$.  At an equlibrium
point all initial conditions are available: $ {\bf X}$ is
the equlibrium point state,  $ \Omega_i = 0 $ and $ y_i = 0 $
at the equlibrium point, and the classical decomposition
furnishes the initial values of $ E $ and $ \dot \Omega_i$. 
Of course, a trajectory started exactly at the equlibrium
will not evolve, but one started nearby will, and its initial
conditions can be calculated from the series expansions about
the equlibrium.  

     Then beginning near an unstable equlibrium, we can integrate
trajectories outward, including the stability exponents and
modal frame coordinates $ y$.  Since the modal transformation
decouples the equations of motion and not the trajectories, an
individual trajectory cuts across different coordinate lines
$ Y$, and with a dense sampling of trajectories the entire
modal frame may be mapped out, starting from the vicinity of
the equlibrium.

     A proof that this procedure 
produces an actual coordinate frame can be sketched as follows.  
The differential equations (\ref{nonlinear}), (\ref{Edot}), 
(\ref{ydot}), (\ref{Omegadot}) and (\ref{Omegadotdot}) 
are just the differential equations we have expanded about
the equlibrium.  When numerically integrated, they
produce unique values of $ {\bf Y}(t)$ and $ {\bf X}(t)$ by
the standard existence theorems for ordinary differential
equations.   Then, at least locally 
$ {\bf X}({\bf Y}) $ is a well defined function, since its  
Jacobian matrix $ \partial {\bf X} / \partial {\bf Y} = E$ is,
by construction, nonsingular.
Then, as an immediate consequence 
$ \partial E_{ij} / \partial y_k = \partial^2 {\bf X}_i /\partial y_j
\partial y_k $ is symmetric with respect to the indices 
$ j$ and $k$, and $ {\bf Y} $ is a coordinate frame.

\section{Separation Near a Periodic Orbit}

     The construction of the standard Floquet solution
(\ref{periodcoef}) begins with the solution to a boundary
value problem to find periodic initial conditions.  In this
process, the variational equations (\ref{varyeqns}) are 
integrated to help find the periodic orbit, and a natural
by--product of this is the monodromy matrix 
$ \Phi (\tau, 0)$, the state transition matrix at one
period.  Then, since the modal vector matrix $ E(t) $ is
periodic, (\ref{periodcoef}) directly shows that the eigenvectors
of the monodromy matrix are the initial modal matrix $ E(0)$,
while the Poincar\'e exponents $ \Lambda_i $ are related to
the eigenvalues $ \lambda_i $ of the monodromy matrix by
$ \Lambda_i = \log \lambda_i / \tau$.  Then the modal
matrix may be propagated for one period using (\ref{Edot})
with initial conditions $ E(0) $ known, and 
$ \omega_i = \Lambda_i$ taken to be constants.

     This must be modified somewhat in the current theory.  
Since the $ E $ matrix forms the basis vectors for a new coordinate
system $ y$, $ E $ must be periodic, and again we have that
the matrix $ E(0) $ is the eigenvector matrix of 
$ \Phi( \tau, 0)$.  But the classical Poincar\'e exponents
cannot be taken to be constant if the determinant of $ E $
must be constant.  Instead, they must be interpreted through
(\ref{exponents}) as the global stability exponents
for the periodic orbit.  This implies the constraint
\begin{equation}
\Lambda_i = {{ \Omega_i (\tau)} \over \tau} ,
\label{Poin2Stab}
\end{equation}
relating the standard Poincar\'e exponent to the exponents
introduced in this work.  
Then, comparing known information to the initial conditions
necessary for time propagation, we still do not know the
initial values of the \it local \rm stability exponents
$ \dot \Omega_i (0)$.

     Taking the expression for $ \ddot \Omega_i $ (\ref{Omegadotdot})
and integrating twice with respect to time along the periodic 
orbit gives
\begin{equation}
\Omega_i (\tau) = \dot \Omega_i (0) \tau + \int_0^{\tau} 
\int_0^{\tau} {{ \partial A_{\alpha \alpha} } \over {\partial x_{\beta}}}
E_{\beta i} E^{-1}_{i \gamma} f_{\gamma} dt^2 .
\end{equation}
Inserting this result into (\ref{Poin2Stab}), the unknown
initial conditions for the local stability exponents are
\begin{equation}
\dot \Omega_i (0) = \Lambda_i - {1 \over { \tau}}
\int_0^{\tau} \int_0^{\tau} {{ \partial A_{\alpha \alpha} } \over 
{\partial x_{\beta}}}
E_{\beta i} E^{-1}_{i \gamma} f_{\gamma} dt^2 .
\label{POics}
\end{equation}
The double integration can be easily performed with the time
propagation algorithm by beginning the integration with
zero initial conditions for $ \dot \Omega_i (0)$.  There is
nothing wrong analytically with (\ref{POics}), and numerically
speaking it sometimes even works.
The difficulty arises when Poincar\'e exponents are far
from zero, leading to numerical problems in inverting an
exponentially growing or decaying $ E $ matrix.  For Hamiltonian
systems this is not necessary, since in view of (\ref{HamdOmdy})
the initial conditions are $ \dot \Omega_i (0) = 0$.

    The above suffices for isolated periodic orbits.  However, for
systems with families of periodic orbit, normal forms for the
matrix $ \dot \Omega $ are required.  Also, some further considerations
come into play in order to construct legitimate coordinate frames
$ {\bf Y}$.  For two dimensional Hamiltonian systems, we may begin 
with the usual normal form
\begin{equation}
\dot \Omega = \left\{ \begin{array}{c c}
 0 & 1 \\
 0 & 0 \end{array} \right\}
\end{equation}
for two dimensional systems.  This form is only possible since
the local stability exponents are zero for Hamiltonian systems.
Then, direct calculation will show that $ \dot \Omega $ commutes
with
\begin{equation}
\Omega = \left\{ \begin{array}{c c}
 0  &  t \\
 0  &  0  \end{array} \right\}
\end{equation}
for any time.  At the end of one period this leads to the
extended eigenvector / eigenvalue problem
\begin{equation}
\Phi( \tau, 0) E - E \left\{ \begin{array}{c c}
 1  & \tau  \\
 0  &  1 \end{array} \right\} = 0.
\end{equation}
As is well known, the regular eigenvector $ {\bf e}_1 $ will be the state
velocity vector $ \dot {\bf X} $.  The extended eigenvector will
then be a solution of $ ( \Phi - I ){\bf e}_2 = \tau {\bf e}_1 $, 
combined with the condition $ {\bf e}_1 \cdot {\bf e}_2 = 0$, 
since the starting point on an adjacent periodic orbit is arbitrary.

     A first integration of the differential equations of the
previous section around the orbit will then confirm that the
matrix $ E $ closes on itself at the end of one period, as it must
if $ {\bf Y}$ is a coordinate frame.  This integration will also
furnish the value of $ y_1 $ at one period.  Since $ {\bf e}_1 (t) $
is the state velocity vector, the new coordinate $ y_1 $ will
be measured along the orbit itself, and herein lies a problem.  
Since there are adjacent periodic orbits, and since the coordinate
$ y_1 $ must have a branch cut, it is necessary to normalize
$ y_1 (\tau)$ to some constant value, in order that $ y_1 $
coordinate space not appear and disappear at the branch cut as
we move from one periodic orbit to an adjacent orbit.  If $ d_1 $
is the multiplicative scale factor for renormalizing $ {\bf e}_1$, 
$ y_1 (\tau) $ is the current maximum value, and say $ 2 \pi $
is the desired maximum value, then
$ d_1 = 2 \pi / y_1 (\tau)$.  Then, to symplectically normalize
the $ E $ matrix, \cite{wie94}, the multiplicative factor $ d_2 $ for
$ {\bf e}_2 $ must be chosen as
$ d_1 d_2 = \left( E^T Z E \right)_{12}$, where $ Z $ is
the usual symplectic matrix.  This done, the renormalized
$ E $ matrix will be symplectic, and the transformation between
$ {\bf X} $ and $ {\bf Y }$ will be canonical.  The branch curve
for the coordinate $ y_1 $ will be the locus of starting points
for the family of periodic orbits, and moving from periodic
orbit to periodic orbit, the $ y_2 $ coordinate obeys
\begin{equation}
 {{ \partial {\bf Y} } \over { \partial {\bf X} }} = E^{-1},
\end{equation}
which is found by differentiating (\ref{coordseries}).  This may
be integrated to track the $ y_2 $ coordinate.

     The non--canonical case is somewhat more difficult, since
the diagonal $ \dot \Omega_{ii} $ are not necessarily zero.
Again assuming a second order system, the diagonal entries are
given by (\ref{dOmdy}).  For a second order system, we write
\begin{equation}
\dot \Omega = \left\{ \begin{array}{c c}
\dot \Omega_{11} & \dot \Omega_{12} \\
0 & \dot \Omega_{22} \end{array} \right\} , \;\;\;
\Omega = \int_0^t \dot \Omega dt = \left\{ \begin{array}{c c}
\Omega_{11} & \Omega_{12} \\
0 & \Omega_{22} \end{array} \right\}  .
\end{equation}
Then, we must ensure that $ \dot \Omega $ and $ \Omega $
commute in order that (\ref{mydcmp}) is valid.  Direct calculation of 
$ \dot \Omega \Omega - \Omega \dot \Omega $ shows that this
occurs if
\begin{equation}
\left( \Omega_{11} - \Omega_{22} \right) \dot \Omega_{12} = 
\left( \dot \Omega_{11} - \dot \Omega_{22} \right) \Omega_{12}
\end{equation}
The existence of such normal forms remains a current research
topic.

\section{Numerical Experiments}

     One system we have studied numerically is Van der Pol's
equation
\begin{equation}
\dot x_1 = x_2, \;\;\; \dot x_2 = - \epsilon ( x^2 - 1 ) \dot x - x 
\end{equation}
for $ \epsilon = 1$.  The parameters of the equlibrium point
decomposition are listed in Table I.  The origin is an unstable spiral
point, and there is the usual stable limit cycle.  This system
makes it possible to do numerical experiments on both the equlibrium
and periodic orbit.

\begin{figure}
\epsfxsize=3.4in
\epsfbox{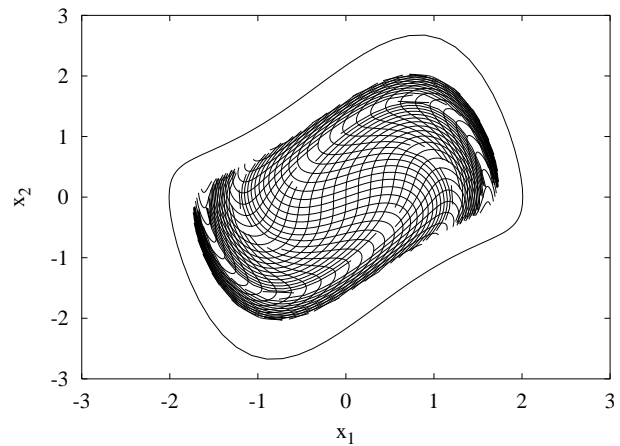}
\vspace{.1in}
\caption{Contours of $ \Re y_1 $ and $ \Im y_1 $ shown on
the $ X $ plane for Van der Pol's equation. \label{vdpycontour}}
\end{figure}

   By integrating numerous trajectories outward from the vicinity
of the equlibrium, and keeping track of the $ y_i $ coordinate
values crossed, Fig. \ref{vdpycontour} is produced.  It shows 
the $ y $ frame coordinate
grid with a contour separation of 0.1, and since the two
coordinates $ y_1 $ and $ y_2 $ are always complex conjugate
for real valued $ x$, we have shown contours of 
$ \Re y_1 $ and $ \Im y_1$.  The limit cycle is shown on the outside
for comparison.  [Some breaks in the contours are numerical
artefacts.]  The origin of the $ y $
frame is at the equlibrium, and it can be seen that the coordinate
frame is collapsing as the limit cycle is approached.  This is, 
however, done maintaining a constant determinant for the $ E $ matrix.
Apparently any coordinate frame tied to the limit cycle will not
match smoothly with the modal coordinate frame tied to the
equlibrium point at the origin.

\begin{figure}
\epsfxsize=3.4in 
\epsffile{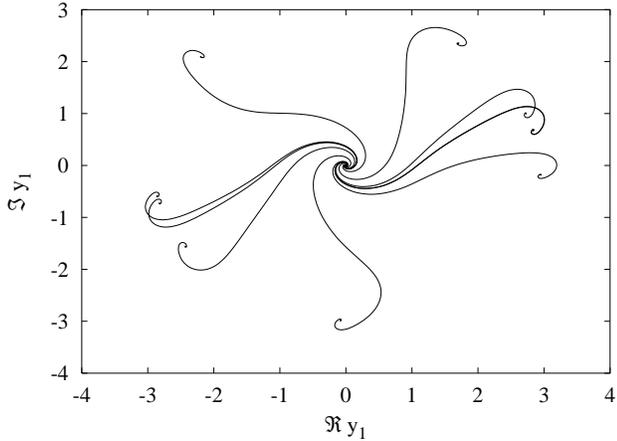}
\vspace{.1in}
\caption{Van der Pol trajectories in the $ \Re y_1 $, $ \Im y_1$
plane. \label{vdpytraj}}
\end{figure}

    Another interesting way to show the trajectories is to plot
them in the $ y $ frame itself, Fig. \ref{vdpytraj}.  Since 
the plot again shows
the real part $ \Re y_1 $ versus the imaginary part $ \Im y_1 $,
the trajectories should not really cross each other, even in the
$ y $ frame.  To the extent that they do, we have pushed the 
numerical calculation too far in an attempt to discover what
the limit cycle maps to in the new coordinate frame.  The problem
occurs because the individual vectors of the $ E $ matrix grow at
nearly the correct rate that the $ y_i $ become constant as the
limit cycle is approached.

\begin{figure}
\epsfxsize=3.4in \epsfbox{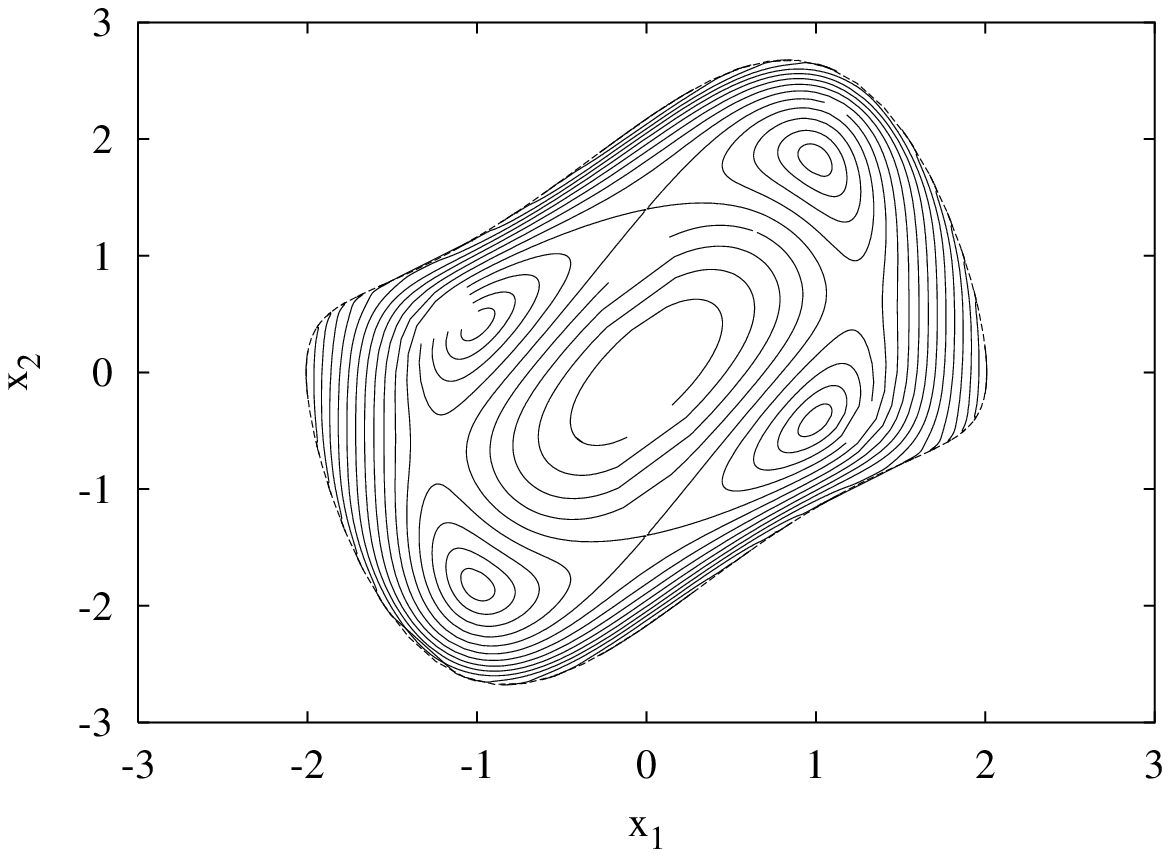}
\vspace{.1in}
\caption{Contour plot of $ \left| \dot \Omega \right| $ on
the $ {\bf X}$ plane. \label{vdpjdotmag}}
\end{figure}

     A similar technique can be used to examine the behavior
of the local stability exponents $ \dot \Omega_i ( {\bf X}$.  
Sampling trajectories eminating from the equlibrium point, 
we have kept track of where certain values of 
$ \left| \dot \Omega_i \right| $ are crossed, and then
combined these into contours.  Figure \ref{vdpjdotmag} shows
the result, with a contour interval of 0.1.  Near the equlibrium
we have $ \dot \Omega_i = 0.5 \pm i \sqrt{3} / 2$, the
value from the constant coefficient system.  The two local
exponents remain complex conjugate across the $ {\bf X}$
plane.  The plot terminates on the limit cycle, since no
trajectories from the equlibrium can cross this line.

\begin{figure}
\epsfxsize = 3.4in
\epsfbox{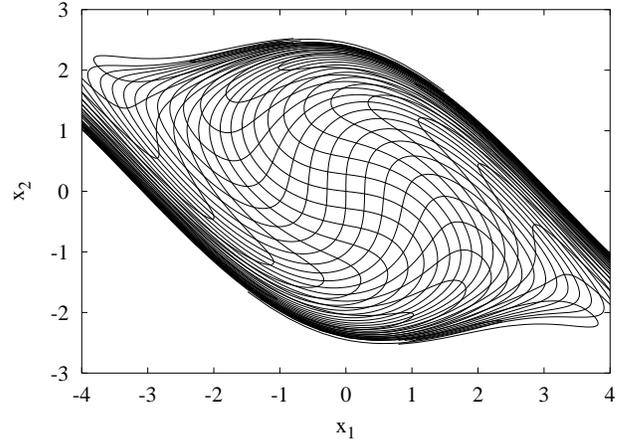}
\vspace{.1in}
\caption{Contours of $ \Re y_1 $ and $ \Im y_1 $ for the
damped pendulum. \label{dmpend25}}
\end{figure}

     The equlibrium point decomposition fails when stability
exponents occur as positive / negative pairs, including centers,
saddles, and in particular all Hamiltonian equlibrium points.  We 
have used the damped pendulum 
\begin{equation}
\dot x_1 = x_2, \;\;\; \dot x_2 =  - c \dot x - \sin x = 0
\end{equation}
as a way to explore this problem.  Figure \ref{dmpend25}
shows the $ y $ frame [with a contour interval of 0.2] 
for the pendulum with a damping factor
of $ c = 0.25$.  The $ y $ frame is relatively flat near the
equlibrium point, but becomes severely distorted as the 
separatrices are approached.   As the damping factor is
decreased towards zero, this twisting of the coordinate
frame becomes more and more severe near the origin, and the
zone of validity of the linearization about the equlibrium
becomes smaller and smaller.  Then, as $ c \rightarrow 0$, 
the twist of the coordinate frame tied to the equlibrium point
becomes infinite.  However, this is just what is observed
numerically with (\ref{order3}) [but not (\ref{order2})] as
the damping vanishes.  Apparently this is structural, since
it occurs for any choice of local stability exponents that
the author has investigated.  So, it appears to be impossible
to use the equlibrium point decomposition to define a global
coordinate frame when the equlibrium is a center or a saddle.

\begin{figure}
\epsfxsize=3.4in
\epsfbox{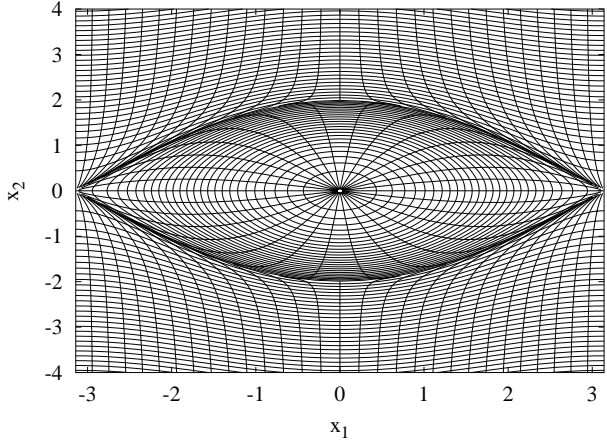}
\vspace{.1in}
\caption{Three $y$ coordinate frames based on the periodic
orbits of the undamped pendulum. \label{pendpo}}
\end{figure}

     This does not mean that coordinate decompositions do not exist
for the undamped pendulum.  Figure \ref{pendpo} shows three $ y $ frames
based on periodic orbits which are canonical, and between them
cover the entire phase plane.  The $ y_1 $ directions are
along the periodic orbits.  The construction was made with the
maximum value of $ y_1 $ as $ 2 \pi$, since clearly $ y_1 $
topologically looks like an angle near the stable equlibrium.  
The stable equlibrium is obviously a singular point for this coordinate
frame, and furthermore the saddle point at $ x = \pm \pi $ is also a
singular point for all three $ y$ coordinate frames.

    We have sought numerical confirmation that using the
methods of section IV actually produces a
decoupling coordinate transformation.  In particular, we have
attempted to calculate the matrix $ \partial g_i / \partial y_j $,
the gradient vector $ \partial \left| E \right| / \partial y_i$ and
$ \partial E_{ij} / \partial y_k$.
The first should be diagonal, with non--zero entries
$ \partial g_i / \partial y_i = \dot \Omega_i$, the second 
should be identically zero, while the last 
quantity should be symmetric in $ j$ and $ k$ if $  Y $ is a
true coordinate frame.  To calculate a numerical partial
derivative, we have used Lagrangian five point numerical
derivatives.  Each numerical partial requires integrating a sheaf of five
trajectories from the vicinity of the equlibrium to straddle
the current point.  Since it is common to use
equal spacing in these points, we have solved a boundary value
problem to find initial $ x $ values that produce equal spacing
in the $ y_i $ coordinate.  As we have chosen to make these
displacements real, this means that it is necessary to integrate
the physical system off of the real $ x $ axis.  Explicitly, 
we have plotted the quantities
\begin{equation}
a = \left| \left| E(t) \right| - \left| E(t_0) \right| \right|,
\end{equation}
the error in propagating the determinant of $ E$, the
maximum deviation of its gradient from zero,
\begin{equation}
b = {\rm Max}_i \left| {{ \partial \left| E \right| }
\over { \partial y_i }} \right| , \nonumber \\
\end{equation}
the maximum error in the local diagonalization of the
equations of motion
\begin{equation}
c = {\rm Max}_{ij} \left| {{ \partial g_i } \over {\partial y_j}}
- \dot \Omega_i \delta_{ij} \right|  ,
\end{equation}
and the maximum violation of the symmetry condition for
a coordinate frame
\begin{equation}
d = {\rm Max}_{jk} \left|  {{ \partial E_{ij} } \over
{ \partial y_k }} - {{ \partial E_{ik} } \over {\partial y_j}}
\right|
\end{equation}
as functions of time.  Since these integrations extend from a
point very close to the equlibrium well into the nonlinear
regieme, and since solution of a boundary value
problem is required to obtain the points necessary for the 
numerical partial derivatives, the author feels that some
numerical error growth is inevitable.

\begin{figure}
\epsfxsize=3.4in \epsfbox{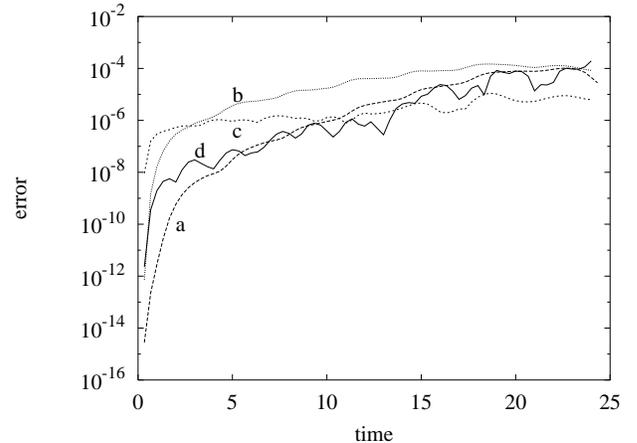}
\vspace{.1in}
\caption{Numerical checks of the decoupling coordinate frame 
    for the damped pendulum. \label{penderror}}
\end{figure}

\begin{figure}
\epsfxsize=3.4in \epsfbox{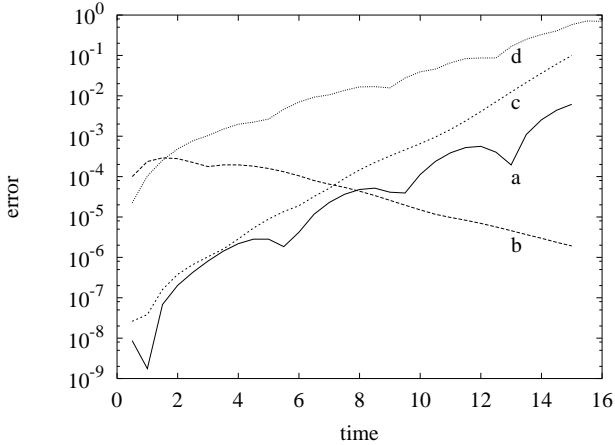}
\vspace{.1in}
\caption{Numerical checks of the decoupling coordinate frame 
    for Van der Pol's equation. \label{vdperror}}
\end{figure}

     Figure \ref{penderror} shows the result of one such calculation
for the damped pendulum with $ c=0.25$, a starting position within
0.001 radian of the origin, and whose ending position is well over
4 radians.  After some initial error growth, all of these tests
indicate that the time propagation method is indeed producing
a true coordinate frame $ {\bf Y} $ which separates the equations
of motion.  Similar results are shown in Fig. \ref{vdperror} for
the Van der Pol system.  Again the starting point is within
0.0001 units of the equlibrium, and the final point is quite
close to the limit cycle.  In this case error growth in the
calculation of these four quantities is more pronounced than
in the case of the pendulum.  But the results confirm that
the $ y_i $ do decouple the equations of motion, and that
the integration is conserving the determinant $ \left| E \right|$.  
Confirmation of the symmetry condition is the
worst numerical verification, since this quantity is the difference of
two numerical partial derivatives.
There is good theoretical reason to believe that each of
these conditions are met.

\section{Discussion and Conclusions}

     In this paper we have shown that the problem of local
stability exponents which integrate along a trajectory to give
global stability exponents is fully equivalent to the problem
of dynamical separation of variables.  The choice of constant
determinant for the modal matrix $ E $ seems compelling for
dissipative systems, and is the only choice permitted for
Hamiltonian systems.  The local stability exponents introduced
by this choice are invariant to any linear change of variables.  
This invariance is less than the total invariance familiar
from constant coefficient linear systems and time-periodic
systems, but is still broader than the degree of invariance
permitted by standard Lyapunov exponents.  We have presented
methods which lead to decoupling transformations in the
vicinity of equlibrium points, periodic orbits, and along
general trajectories eminating from equlibrium points.  
Such decoupling transformations may not exist in the vicinity
of center and saddle point equlibria when the stability
exponents of the equlibrium exist as positive/negative pairs.  
However, in this case it appears possible to base decoupling
coordinate transformations on the surrounding periodic orbits.

     Much work remains to be done.  We are currently exploring
control applications, as well as continuing work on decoupling
transformations near periodic orbits.  Another area of interest
is the relation of this method to the ``geometrodynamics''
approach, which attempts to decouple the \it trajectories \rm 
of a dynamical system, and not necessarily the underlying
equations of motion.  Our method is based on decoupling the
equations of motion, but seems to become a trajectory--based
decoupling for Hamiltonian systems.

\section{Appendix A.}

     The form of the separation conditions near the equlibrium 
suffer from the fact that the ``columns'' of the partial
derivatives of $ E $ are coupled.  We have used equations
(\ref{order2}) and (\ref{order3}) for numerical checks, and
have employed an alternate form for the solution that is
more efficient numerically.

     Instead of beginning with the separation conditions, begin
with the modal matrix equation of motion (\ref{Edot}), written as
\begin{equation}
\left( A_{j \alpha} - \dot \Omega_i \delta_{j \alpha} \right)
E_{\alpha i} = {{\partial E_{ji} } \over {\partial y_{\beta} }}
E^{-1}_{\beta \alpha} f_{\alpha} .
\end{equation}
Evaluation at the equlibrium point immediately yields the eigenvalue
/ eigenvector problem for the first order terms.  Then, taking
a partial derivative, evaluating at the equlibrium, 
and simplifying there results
\begin{eqnarray}
 A_{j \alpha} {{ \partial E_{\alpha i} } \over {\partial y_{k}}}
&-& \dot \Omega_i {{ \partial E_{ji}} \over {\partial y_{k} }}
- \dot \Omega_{k} {{\partial E_{ji}} \over {\partial y_{k} }}
\nonumber \\
&-& {{\partial \dot \Omega_i } \over {\partial y_i }} \delta_{k i}
E_{ji} = - {{\partial A_{j \alpha}} \over {\partial x_{\gamma} }}
E_{\gamma k} E_{\alpha i} .
\end{eqnarray}
The advantage of this form is that each ``column'' (e.g. index
$ i $) is decoupled from all the others, reducing the
order of the linear system by a factor of $ N $, the order
of the dynamical system.  Treating the quantities 
$ \partial \dot \Omega_i /\partial y_i $ as known, we must
solve $ N $ systems of $ N^2 $ linear equations, not one
system of order $ N^3$.
Continuing to the third order gives
\begin{eqnarray}
& & A_{j \alpha} {{\partial^2 E_{\alpha i}} \over { \partial y_{k}
\partial y_{l} }} - \left ( \dot \Omega_i + \dot \Omega_{k}
+ \dot \Omega_{l} \right) {{ \partial^2 E_{ji} } \over {
\partial y_{k} \partial y_{l} }} \nonumber \\
&-& {{\partial^2 \dot \Omega_i } \over { \partial y_i^2 }} E_{ji}
\delta_{k i} \delta_{l i} = 
- {{\partial^2 A_{j \alpha}} \over {\partial x_{\delta} \partial 
x_{\epsilon}}}
E_{\epsilon l} E_{\delta k} E_{\alpha i} \nonumber \\
&-& 
{{\partial A_{j \alpha}} \over {\partial x_{\delta}}} \left(
{{\partial E_{\alpha i}} \over {\partial y_{k}}} E_{\delta l} +
{{\partial E_{\delta k}} \over {\partial y_{l}}} E_{\alpha i} +
{{\partial E_{\alpha i}} \over {\partial y_{l}}} E_{\delta k}
\right) \nonumber \\
&+& 
{{\partial \dot \Omega_i } \over {\partial y_i}} \delta_{i l}
{{\partial E_{ji}} \over {\partial y_{k}}} +
{{\partial \dot \Omega_i } \over {\partial y_i}} \delta_{k i}
{{\partial E_{ji}} \over {\partial y_{l}}}  \\
&-& 
{{\partial E_{ji}} \over {\partial y_{\tau}}} E^{-1}_{\tau \delta}
{{\partial E_{\delta l}} \over {\partial y_{k}}}
\dot \Omega_{l} - 
{{ \partial E_{ji} } \over { \partial  y_{\tau}  }}  E^{-1}_{\tau \delta} 
{{ \partial E_{\delta k} } \over { \partial y_l }}
\dot \Omega_k \nonumber \\
&+&
{{ \partial E_{ji} } \over { \partial y_{\tau} }} E^{-1}_{\tau \alpha}
A_{\alpha \gamma} {{\partial E_{\gamma k} } \over { \partial y_l }}
+
{{\partial E_{ji} } \over { \partial y_{\beta} }} E^{-1}_{\beta \alpha}
{{ \partial A_{\alpha \gamma} } \over { \partial x_{\tau} }}
 E_{\gamma k} E_{\tau l}  . \nonumber
\end{eqnarray}
This form also has the ``column'' separation property, since
all of the second partials of $ E $ with second index $ i $ can be
solved for at once.

\begin{center}
Table I. \\
Van der Pol Equlibrium Decomposition
\end{center}
\begin{tabular}{l  l}
Order 1 \\
\hline \hline
  $\dot \Omega_1$ &   $ 0.50000000000000 +   0.86602540378444 i $ \\
  $ e_{11}$ & $  0.70710678118655 +  0. i $ \\
  $ e_{21}$ & $  0.35355339059327 +  0.61237243569579 i $ \\
  $\dot \Omega_2$ & $   0.50000000000000  -0.86602540378444 i $ \\ 
  $ e_{12}$ & $   0.70710678118655 + 0. i $ \\
  $ e_{22}$ & $   0.35355339059327  -0.61237243569579 i $ \\
\hline
Order 2 \\
\hline \hline
  $\dot \Omega_{11} $ & $  0.+  0.i $ \\
  $ e_{111} $ & $   0.+  0.i $ \\
  $ e_{112} $ & $   0.+  0.i $ \\
  $ e_{211} $ & $   0.+  0.i $ \\
  $ e_{212} $ & $   0.+  0.i $ \\
  $ \dot \Omega_{22} $ & $  0.+  0.i $ \\
  $ e_{121}$ & $    0.+  0.i $ \\
  $ e_{122}$ & $    0.+  0.i $ \\
  $ e_{221}$ & $    0.+  0.i $ \\
  $ e_{222}$ & $    0.+  0.i $ \\
\hline
Order 3 \\
\hline \hline
$\dot \Omega_{111}$ & $  -1.00000000000000+  0.i $ \\
$ e_{1111}$ & $ 0.0951874513135   -0.0235527859883 i $ \\
$ e_{1112}$ & $  -0.53033008588991+   0.30618621784790 i $ \\
$ e_{1121}$ & $  -0.53033008588991+   0.30618621784790 i $ \\
$ e_{1122}$ & $  -0.53033008588991   -0.30618621784790 i $ \\
$ e_{2111}$ & $  -0.50313367122889+   0.21197507389470 i $ \\
$ e_{2112}$ & $   -1.0606601717798  + 0. i $ \\ 
$ e_{2121}$ & $   -1.0606601717798 + 0. i $ \\
$ e_{2122}$ & $   -1.0606601717798 + 0. i $ \\
$\dot \Omega_{222}$ & $  -1.00000000000000+  0. i $ \\
$ e_{1211}$ & $  -0.53033008588991+   0.30618621784790 i $ \\
$ e_{1212}$ & $  -0.53033008588991  -0.30618621784790 i $ \\
$ e_{1221}$ & $  -0.53033008588991  -0.30618621784790 i $ \\
$ e_{1222}$ & $  0.0951874513135 +  0.0235527859883 i $ \\
$ e_{2211}$ & $   -1.0606601717798   + 0. i $ \\
$ e_{2212}$ & $   -1.0606601717798 + 0. i $ \\
$ e_{2221}$ & $   -1.0606601717798 + 0. i $ \\
$ e_{2222}$ & $   -0.50313367122889  -0.21197507389470 i $ \\
\hline
\end{tabular}

\end{document}